\documentclass[final,5p,times,twocolumn,authoryear]{elsarticle}
\usepackage{amssymb}
\usepackage{lipsum}
\usepackage{amsmath}
\usepackage{cuted}
\usepackage{graphicx}    
\usepackage{subcaption}  
\usepackage{float}

\journal{Physics Letters B}
\makeatletter
\def\ps@pprintTitle{%
\let\@oddhead\@empty
\let\@evenhead\@empty
\let\@oddfoot\@empty
\let\@evenfoot\@oddfoot}
\makeatother

\begin{document}

\begin{frontmatter}



\title{Neutrino phenomenology and Dark matter in a left-right asymmetric model with non-holomorphic modular $A_{4}$ group}

\author[first]{Bhabana Kumar\fnref{fn1}}
\author[first]{Mrinal Kumar Das\fnref{fn2}}

\affiliation[first]{
  organization={Department of Physics, Tezpur University},
  city={Tezpur},
  postcode={784028}, 
  state={Assam},
  country={India}
}

\fntext[fn1]{Email: bhabana12@tezu.ernet.in}
\fntext[fn2]{Email: mkdas@tezu.ernet.in}

\begin{abstract}
We present a model constructed within a non-supersymmetric framework capable of explaining both current neutrino oscillation data and the observed dark matter relic abundance. In this study, the Yukawa couplings are expressed as polyharmonic Maa\ss{} forms, and the non-supersymmetric left-right symmetric model is realized through the $\Gamma_{3}$ modular group, with neutrino masses generated via the Type II seesaw dominance mechanism. The analysis focuses on determining the neutrino oscillation parameters, the effective Majorana mass arising from the standard contribution, and the dark matter relic density. Our results indicate that the model strongly favours the normal mass hierarchy over the inverted one and prefers the lower octant for the mixing angle $\theta_{23}$. Furthermore, the effective Majorana mass is predicted to lie in the range $10^{-3}\,\text{eV}$ to $0.1\,\text{eV}$. In addition, the lightest sterile neutrino present in the model is considered a viable dark matter candidate. A sterile neutrino mass in the range $10~\text{keV}$ to $30~\text{keV}$ is found to yield consistent results for both the relic density and active-sterile mixing angles.
\end{abstract}


\end{frontmatter}

%\tableofcontents



\section{Introduction}
\label{introduction}
The discovery of neutrino oscillations shows that the Standard Model (SM) of particle physics is not a complete theory. In the SM, neutrinos are considered massless and electrically neutral Dirac fermions. However, the observation of neutrino oscillations implies that neutrinos must possess small, non-zero, and non-degenerate masses. Over the years, various neutrino oscillation experiments have successfully measured the three mixing angles and two mass-squared differences. Still, several important questions remain unanswered, such as whether neutrinos are Dirac or Majorana particles, the type of mass ordering (normal or inverted), the octant of $\theta_{23}$, and the exact value of the CP-violating phase $\delta_{CP}$. These open questions make the study of neutrinos highly interesting.\\
One of the most intriguing processes that can shed light on the nature of neutrinos is neutrinoless double beta ($0\nu\beta\beta$) decay \cite{Schechter:1981bd,Schwingenheuer:2012jt}. This process involves a nucleus undergoing two successive beta decays without emitting neutrinos in the final state. The observation of $0\nu\beta\beta$ decay would provide definitive proof of the Majorana nature of neutrinos. Ongoing experimental searches for this decay process also provide upper bounds on the effective Majorana mass and lower bounds on the half-life of the decay. Current results indicate that the effective Majorana mass should be below $0.165$--$0.065$~eV, while the half-life must exceed $10^{26}$~years.\\
In this work, we calculate the effective Majorana mass from the exchange of light neutrinos with left-handed currents. The formula for the standard contribution is given by
\begin{equation}\label{W4Q1}
    m^{\nu}_{ee,L} = \sum^{3}_{i=1} U^{\nu\nu^{2}}_{ei}\, m_{\nu_{i}} \, ,
\end{equation}
%%%%%%%%%%%%%%%%%%%%%%%%%%%%%%%%%%%%%%%%%%%%%%%%%%%%%%%%%%%%%%%%%%%%%%%%%
Not only does the SM fail to explain neutrino oscillations, but it also cannot account for one of the most compelling open questions in modern physics, the existence of dark matter. Observations suggest that approximately $26.8\%$ of the Universe is composed of dark matter. According to Planck data\cite{Planck:2015fie}, the relic abundance of dark matter is given by  
\begin{equation}
    \Omega_{DM}h^{2} = 0.1187 \pm 0.0017.
\end{equation}

Although there is substantial evidence supporting the existence of dark matter, the nature of the dark matter particle remains unknown. There are many studies that have proposed keV-scale sterile neutrinos as dark matter candidates\cite{Drewes:2016upu,Merle:2017jfn,Dodelson:1993je}. In this work, we consider a keV-scale sterile neutrino as a potential warm dark matter (WDM) candidate. A sterile neutrino of this scale can decay into a SM neutrino via the channel $S \rightarrow \nu_{l} + \gamma$, producing a monoenergetic X-ray line with energy $E_{\gamma} = \frac{m_{S}}{2}$. 

Searches for such X-ray signals have been conducted by various space-based observatories, which have imposed strong constraints on the mass and mixing angle parameter space of sterile neutrinos. Sterile neutrinos can be produced in the early Universe through their small mixing with active neutrinos, a process known as the Dodelson--Widrow (DW) mechanism. The mixing angle between sterile and active neutrinos is typically very small, of the order $10^{-10}$ or less. The relic abundance and decay width of sterile neutrinos depend on the mixing angle $\sin^{2}2\theta_{DM}$ and their mass $m_{S}$, as described by the following relations:
\begin{equation}\label{W4Q2}
\begin{aligned}
  \Omega_{DM}h^{2} \approx & \; 1.1 \times 10^{7} 
  \sum C_{\alpha}(m_{S}) |U_{\alpha S}|^{2}
  \left(\frac{m_{S}}{\text{keV}}\right)^{2} \\
  \approx & \; 0.3 
  \left(\frac{\sin^{2}2\theta_{DM}}{10^{-10}}\right)
  \left(\frac{m_{S}}{\text{keV}}\right)^{2}.
\end{aligned}
\end{equation}

The decay width of the sterile neutrino into a SM neutrino is given by\cite{Bezrukov:2009th,Boruah:2022csq}
\begin{equation}\label{W4Q3}
  \Gamma = 1.38 \times 10^{-32} 
  \left(\frac{\sin^{2}2\theta}{10^{-10}}\right)
  \left(\frac{m_{S}}{\text{keV}}\right)^{5} \, \text{s}^{-1},
\end{equation}
where $\sin^{2}2\theta_{DM} = 4 \sum_{\alpha=e,\mu,\tau} |U_{\alpha S}|^{2}$ and $U_{\alpha S} = M_{D} M^{-1} U_{S}$.\\
%%%%%%%%%%%%%%%%%%%%%%%%%%%%%%%%%%%%%%%%%%%%%%
We consider a non-supersymmetric left-right asymmetric model\cite{Chang:1984uy,Chang:1983fu,Senapati:2020alx,Awasthi:2013ff,Parida:2012sq} to explain the above mentioned phenomena. In this framework, one sterile neutrino is introduced per generation, with the lightest sterile neutrino serving as the dark matter candidate. The neutrino mass is generated through the Type~II seesaw dominance mechanism, for which an additional scalar doublet is included in the model.  

The left-right asymmetric model is realized using the $\Gamma_{3}$ modular group. A key advantage of employing modular symmetry in model building is that it eliminates the need for introducing extra scalar fields, known as flavons, to break the flavor symmetry. Instead, in such models, the modulus $\tau$ itself is responsible for breaking the flavor symmetry. This feature makes modular symmetry more attractive compared to conventional discrete symmetries.  

 Modular symmetry is associated with supersymmetric frameworks, where the Yukawa couplings are required to be holomorphic functions of the modulus $\tau$. However, holomorphicity is not a necessary condition in non-supersymmetric framework. In our model, we have considered the non-supersymmetric framework, where the Yukawa couplings correspond to polyharmonic Maa\ss{} forms, which contain both holomorphic and non-holomorphic components. The idea of employing non-holomorphic modular symmetry in model building has been discussed in~\cite{Qu:2024rns,Nomura:2024atp,Ding:2024inn,Nomura:2024vzw,Li:2024svh,Nomura:2024nwh,Kumar:2024uxn,Okada:2025jjo,Kobayashi:2025hnc,Qu:2025ddz,Kumar:2025bfe}.\\
 %------------------------------------------
The present study focuses on constructing a neutrino mass model capable of addressing unresolved issues in the neutrino sector. The framework incorporates a keV-scale sterile neutrino, which serves as a viable dark matter candidate. The model favours the normal Hierarchy (NH) of neutrino masses over the inverted Hierarchy (IH) and predicts the atmospheric mixing angle $\theta_{23}$ to lie in the lower octant. Furthermore, it yields satisfactory results for the active-sterile mixing angle, relic density, and decay width of the dark matter candidate into SM neutrinos. Overall, the proposed framework provides a minimal yet comprehensive approach that simultaneously explains neutrino masses and mixings while accommodating a dark matter candidate.\\
%-------------------------------------------
The paper is organized as follows. In Section~\ref{W4S1}, we briefly discuss the left-right symmetric model and the Type~II seesaw dominance mechanism. Sections~\ref{W4S2} and \ref{W4S3} are devoted to the description of the model and the numerical analysis, respectively. In Section~\ref{W4S4}, we present the results of our study, and finally, in Section~\ref{W4S5}, we summarize our findings in the conclusion.
%-------------------------------------------------------------------------------------------------------------------------------------------------------------------------------
\section{\textbf{Left right symmetric model with Extended Type II seesaw mechanism}}\label{W4S1}
The left-right symmetric model is based on the gauge group $SU(2)_{L}\times SU(2)_{R}\times U(1)_{B-L}\times SU(3)_{C}~(\mathcal{G}_{2213})$. The particle contents and their charge assignments under the gauge group $\mathcal{G}_{2213}$ of the minimal LRSM are given below
\begin{align*}
    L_{L} &= \begin{pmatrix}
        \nu_{L} \\
        e_{L}
    \end{pmatrix} \sim (2,1,-1,1), \hspace{1cm}
    L_{R} = \begin{pmatrix}
        \nu_{R}\\
        e_{R}
    \end{pmatrix} \sim (1,2,-1,1) \\[10pt]
    q &= \begin{pmatrix}
        u_{L}\\
        d_{L}
    \end{pmatrix} \sim (2,1,\tfrac{1}{3},3), \hspace{1cm}
    q_{R} = \begin{pmatrix}
        u_{R} \\
        d_{R}
    \end{pmatrix} \sim (1,2,\frac{1}{3},3)
\end{align*}
%%%%%%%%%%%%%%%%%%%%%%%%%%%%%%%%%%%%%%%%%%%%%%%%%%%%%%%%%%%%%%%%%%%%%%%%%%%%%
Now, if the scalar sector of the model contains only a doublet, then it is possible to get Dirac masses for neutrinos; however, to get the Majorana masses for neutrinos, it depends on how the LRSM gauge group breaks down to the SM gauge group. There are three possible ways by which the LRSM gauge group can break down to the SM gauge group. 
\begin{itemize}
    \item Using scalar doublets $\chi_{L, R}$, in which the right-handed scalar doublet $\chi_{R}$ is responsible for breaking the left-right symmetry. In this scenario, the neutrino mass is Dirac in nature. 
    \item  Using scalar triplets $\Delta_{L,R}$. In this case, the LRSM gauge group is broken down to the SM gauge group when the scalar triplet $\Delta_{L,R}$ gets a non-zero vacuum expectation value (VEV). The scalar triplet $\Delta_{R}$ carries a $B-L$ charge which is equal to $2$, and in this case, the neutrino masses obtained are Majorana in nature and violate the lepton number by two units.
    \item Using both scalar doublet $\chi_{L,R}$ and scalar triplet $\Delta_{L,R}$. 
\end{itemize}
Usually in LRSM, the neutrino masses are generated by giving a non-zero VEV to the scalar triplet and the bidoublet. The Yukawa interaction term can be written as 
\begin{equation}\label{W4Q4}
-\mathcal{L} = \overline{l_{L}^{c}}( Y_1 \Phi + Y_2 \tilde{\Phi}) \l_{R} + \frac{1}{2} f_{ij} ( \overline{(l_{L_{i}})^c} l_{L_{j}} \Delta_L + \overline{(l_{R_i})^c} l_{R_j} \Delta_R) + \text{h.c.}
\end{equation}
The first two terms in the Lagrangian give the Dirac mass matrix $M_{D}$ and the last two terms give the mass matrix $M_{L}$ and $M_{R}$ due to the non zero VEV of the scalar triplet $\Delta_{L}$ and $\Delta_{R}$ respectively. After the spontaneous symmetry breaking of the Lagrangian given in equation \eqref{W4Q4} results in a $6\times 6$ mass matrix. In the basis of $(\nu_{L},N^{c}_{R})$ this mass matrix can be written in the following way
\begin{equation}\label{W3Q2}
    \mathbf{M}=\begin{pmatrix}
                   M_{L} & M_{D} \\
                   M^{T}_{D} & M_{R}
             \end{pmatrix}
\end{equation}
The light neutrino mass matrix can be obtained by diagonalizing the above matrix, and in this case, the obtained light neutrino mass matrix is a combination of both Type I and Type II formulas. 
\begin{equation}\label{W4Q5}
    m_{\nu}= -M_{D}M^{-1}_{R}M^{T}_{D} + M_{L}
\end{equation}
The first term in the equation \eqref{W4Q5} represents the contribution due to Type I seesaw, and the second term represents the contribution due to Type II seesaw. 
In the present work, we are interested in exploring different beyond SM phenomena by considering that neutrino mass generation is dominated by the Type II seesaw dominance \cite{Pritimita:2016fgr}. Also we have constructed the model using the gauge group $SU(3)_{C}\times SU(2)_{L}\times U(1)_{R}\times U(1)_{B-L}~ (\mathcal{G}_{3211})$ in which the $SU(2)_{R}$ gauge group of LRSM is break down to $U(1)_{R}$ by giving a non zero VEV to a scalar triplet $\Sigma$. So within the considered model, the right-handed charged leptons and right-handed neutrinos no longer transform as a doublet. The neutrino mass is generated via the extended Type II seesaw mechanism, and for this purpose, we have added one sterile fermion $S_{i}$ per generation. Also, the scalar sector of the model contains a bidoublet, a scalar triplet and also a scalar doublet. The particle contains of the model and its charge assignments under the considered gauge group is given in Table \ref{W4T1}. 
\begin{table}[ht]
	\centering
	\begin{tabular}{|c|c|c|c|c|c|c|c|c|c|}
		\hline
		
		Field & $l_{R_{i}}$ & $l_{L_{i}}$ & $N_{R}$ & $S$ & $\Phi$ & $\Delta_{R}$ & $\chi_{R}$ & $\Delta_{L}$&$\chi_{L}$ \\ \hline
		
		$SU(2)_L$ & 1 & 2 & 1 & 1 & 2 & 1 & 1 & 3& 2 \\ \hline
		$U(1)_R$ & $-\frac{1}{2}$ & 0 & $\frac{1}{2}$ & 0 & $\frac{1}{2}$ & $-1$ & $\frac{1}{2}$ & 0 & 0 \\ \hline

	\end{tabular}
	\caption{Charge assignment under $SU(2)_L \times U(1)_R \times U(1)_{B-L}$ for the particle content of the model}
	\label{W4T1}
\end{table}
The Yukawa Lagrangian associated with the model is given in equation \eqref{W4Q5}
\begin{equation}\label{W4Q5}
\begin{aligned}
\mathcal{W} =\ & \overline{l_{L}}Y_{l}\Phi l_{R} + f_{l}\overline{l^{C}_{L}}l_{L}\Delta_{L}+f_{R}\overline{N^{C}_{R}}N_{R}\Delta_{R}+F_{R}\overline{N_{R}}\chi_{R}S^{C} \\
 &+ F_{L}\overline{l_{L}}\chi_{L}S + \mu SS + \text{h.c} \\
 &\supset M_{D}\nu_{L}N_{R} + M_{L}\overline{\nu^{C}_{L}}\nu_{L} + M_{R}\overline{N^{C}_{R}}N_{R} + M\overline{N_{R}}S \\
 & + \mu_{L}\overline{\nu^{C}_{L}}S + \mu_{s}S.S
\end{aligned}
\end{equation}
%%%%%%%%%%%%%%%%%%%%%%%%%%%%%%%%%%%%
The term $\mu_{s}SS$ can take any value, and we have considered its value as zero. We have also considered the $<\chi_{L}> \to 0$. After the spontaneous symmetry breaking of the Yukawa Lagrangian given in equation \eqref{W4Q5}, we can write the $9\times 9$ mass matrix in the following way
\begin{equation}\label{W4Q6}
    \mathcal{M}= \begin{pmatrix}
                   M_{L} & 0 & M_{D} \\
                   0 & 0 & M \\
                   M^{T}_{D} & M^{T} & M_{R}
                \end{pmatrix}
\end{equation}
Considering the mass hierarchy $M_{R}> M> M_{D} >> M_{L}$ and after block diagonalization of the matrix $\mathcal{M}$, the mass matrices for the active neutrino, right-handed neutrino and sterile neutrino are given as 
\begin{align}\label{W4Q6}
    m_{\nu} &= M_L, \nonumber \\
    M_N &= M_R = \frac{v_R}{v_L} M_L, \nonumber \\
    M_S &= -M M_R^{-1} M^T
\end{align}
The $9\times 9$ unitary matrix responsible for the complete diagonalization of the matrix given in equation \eqref{W4Q6} is given below

\begin{equation}
    V=\begin{pmatrix}
        U_{\nu} & M_{D}M^{-1} U_{S} & M_{D}M^{-1}_{R}U_{N} \\
        (M_{D}M^{-1})^{\dagger}U_{\nu} & U_{S} & MM^{-1}_{R}U_{N} \\
        0 & -(MM^{-1}_{R})^{\dagger}U_{S} & U_{N}
    \end{pmatrix}
\end{equation}
To obtain the eigenvalues we further diagonalized the  matrices given in equation \eqref{W4Q6} by their respective unitary matrices as follows
\begin{gather} \label{W4Q7}
	\begin{aligned}
		\hat{m}_{\nu} & = U^{\dagger}_{\nu}m_{\nu}U_{\nu}^{*} = diag(m_{\nu_{1}}, m_{\nu_{2}}, m_{\nu_{3}})\\
		%%%%%%%%%%%%%%%%%%%%%%%%%%%%%%%%%%%%%%%%%%%%%%%%%%%%%%%%%%%%
		\hat{m}_{S} & =U^{\dagger}_{S}m_{s}U^{*}_{S} = diag(m_{S_{1}}, m_{S_{2}}, m_{S_{3}}) \\
		%%%%%%%%%%%%%%%%%%%%%%%%%%%%%%%%%%%%%%%%%%%%%%%%%%
		\hat{m}_{R} & = U^{\dagger}_{N}m_{R}U^{*}_{N} = diag(m_{R_{1}}, m_{R_{2}}, m_{R_{3}})~~~. \\
		%%%%%%%%%%%%%%%%%%%%%%%%%%%%%%%%%%%%%%%%%
	\end{aligned}
\end{gather}
Where $U_{\nu}=U_{l}U_{PMNS}$ is the unitary matrix responsible for diagonalizing the light neutrino mass matrix. The matrix $U_{PMNS}$ is a unitary matrix and it can be parametrized by using three mixing angles and three phases, among which one is a Dirac CP phase denoted as $\delta_{CP}$ and two are Majorana phases ($\alpha$, $\beta$). One can represent all the mixing angles in terms of the elements of the $U_{PMNS}$ matrix as given in equation \eqref{W4Q7}, and their $3\sigma$ values of neutrino oscillation parameters are given in \cite{Esteban:2024eli}. We have used those $ 3\sigma$ ranges for our numerical analysis part.
%%%%%%%%%%%%%%%%%%%%%%%%%%%%%%%%%%%%%%%%%%%%%%%%%%%%%%%
\begin{equation}\label{W4Q7}
\begin{aligned}
\sin^{2}\theta_{13} = |(U_{PMNS})_{13}|^{2}, \qquad
\sin^{2}\theta_{23} &= \frac{|(U_{PMNS})_{23}|^{2}}{1-|(U_{PMNS})_{13}|^{2}}, \\[6pt]
\sin^{2}\theta_{12} = \frac{|(U_{PMNS})_{12}|^{2}}{1-|(U_{PMNS})_{13}|^{2}}.
\end{aligned}
\end{equation}

%%%%%%%%%%%%%%%%%%%%%%%%%%%%%%%%%%%%%%%
The Dirac CP phase, Jarlskog invariant and Majorana phases can be determined from $U_{PMNS}$ matrix, and their relations are given in equations \eqref{W4Q8} and \eqref{W4Q9}, respectively
%%%%%%%%%%%%%%%%%%%%%%%%%%%%%%%%%%%%%%%%%%%%%%%%%%%%%%%%%%%

\begin{equation}\label{W4Q8}
	J_{CP}= Im[U_{e1}U_{\mu2}U_{e2}^{*}U_{\mu1}^{*}] = s_{23}c_{23}s_{12}c_{12}s_{13}c^{2}_{13}\sin\delta_{CP} 
\end{equation}
%%%%%%%%%%%%%%%%%%%%%%%%%%%%
\begin{equation}\label{W4Q9}
	Im[U_{e1}^{*}U_{e2}]=c_{12}s_{12}c^{2}_{13}\sin\alpha ~, ~~ Im[U_{e1}^{*}U_{e3}]=c_{12}s_{13}c_{13}\sin(\beta-\delta_{CP}). 
\end{equation}
%%%%%%%%%%%%%%%%%%%%%%%%%%%%%%%%%%%%%%%%%%%%%%%%%%%%%%%%%%%%%%%%%%%%%%%%%%%%%%%%%%%%%%%%%%%%%%%%%%%%%%%%%%%%%%%%%%%%%%%%%%%%%%%%%%%%%%%%%%%%%%
\section{The Model}\label{W4S2}
%\lipsum[1]
In this work, we have realized the model using the $\Gamma_{3}$ modular group, which is isomorphic to the $A_{4}$ group. The light neutrino mass is generated via the type-II seesaw mechanism. The particle content of the model is given in Table~\ref{W4T1}. To construct the model, we consider the left-handed charged leptons to transform as a triplet, while the right-handed charged leptons $L_{R_{i}}$ transform as $1$, $1^{\prime}$, and $1^{\prime \prime}$ under the $A_{4}$ group. The right-handed neutrinos $N_{R}$ transform as $1$, $1^{\prime}$, and $1^{\prime \prime}$, whereas the sterile neutrino $S$ transforms as a triplet under the $A_{4}$ group. All the scalar fields associated with the model transform as trivial singlets under the $A_{4}$ group. The charge assignments under the $A_{4}$ group and the modular weights of all the particles contained in the model are given in Table~\ref{W4T2}. 

\begin{table}[ht]
	\centering
	\begin{tabular}{|c|c|c|c|c|c|c|c|c|c|}
		\hline
		
		Field & $L_{R_{i}}$ & $L_{L_{i}}$ & $N_{R}$ & $S$ & $\Phi$ & $\Delta_{R}$ & $\chi_{R}$ & $\Delta_{L}$&$\chi_{L}$ \\ \hline
		
		$A_{4}$ & $1,1^{\prime},1^{\prime \prime}$ & 3 & $1, 1^{\prime}, 1^{\prime \prime}$ & 3 & 1 & 1 & 1 & 1& 1 \\ \hline
		$k$ & 0 & 0 & 0 & 0 & 0 & 0 & 0 & 0 & 0 \\ \hline
	
	\end{tabular}
	\caption{Charge assignment under $A_{4}$ group for the particle content of the model}
	\label{W4T2}
\end{table}
\begin{table}[ht]
		\centering
		\begin{tabular}{|c|c|c|} \hline
			& Polyharmonic Maa\ss{} ($Y^{k}_{r}$) form\\ \hline
			$A_{4}$ & $3, 1$ \\ \hline
			$k_{I}$ & $0$ \\ \hline 
		\end{tabular}
		\caption {Charge assignment and modular weight for Yukawa coupling}
		\label{W2T3}
	\end{table}
The Yukawa Lagrangian of the model can be divided into two parts: the charged lepton sector $\mathcal{L}_{l}$ and the neutral lepton sector $\mathcal{L}$. The charge assignments under the $A_{4}$ group are chosen in such a way that all the terms in the Yukawa Lagrangian become trivial singlets, i.e., invariant under the $A_{4}$ group.
%%%%%%%%%%%%%%%%%%%%%%%%%%%%%%%%%%%%%%%%%%%%%%%%%%%%%%%%%%%%%%%%%%%%%%%%%%%%%%%%%%%%%%
\subsection{Yukawa Lagrangian for the charged lepton}
The Yukawa Lagrangian for the charged lepton sector associated with the model is given in equation~\eqref{W4Q1}:
\begin{equation}\label{W4Q1}
    \mathcal{L}_{l} = a_{1}\,\Phi\,(l_{L}Y^{0}_{3})_{1}l_{R_1} 
    + a_{2}\,\Phi\,(l_{L}Y^{0}_{3})_{1^{\prime\prime}}l_{R_2} 
    + a_{3}\,\Phi\,(l_{L}Y^{0}_{3})_{1^{\prime}}l_{R_3}.
\end{equation}

From the above Yukawa Lagrangian, the charged lepton mass matrix can be constructed, which is given in Eq.~\eqref{W4Q2}:
\begin{equation}\label{W4Q2}
    M_{l} = v \begin{pmatrix}
        a_{1} Y^{(0)}_{3,1} & a_{2} Y^{(0)}_{3,3} & a_{3} Y^{(0)}_{3,2} \\
        a_{1} Y^{(0)}_{3,3} & a_{2} Y^{(0)}_{3,2} & a_{3} Y^{(0)}_{3,1} \\
        a_{1} Y^{(0)}_{3,2} & a_{2} Y^{(0)}_{3,1} & a_{3} Y^{(0)}_{3,3} 
    \end{pmatrix}.
\end{equation}

The charged lepton mass matrix is not diagonal, and it can be diagonalized by using a bi-unitary transformation,  
\[
\text{diag}(m_{e}, m_{\mu}, m_{\tau}) = V^{\dagger}_{L} M_{l} V_{R},
\]  
which implies that  
\[
V^{\dagger}_{L} M_{l} M_{l}^{\dagger} V_{L} = \text{diag}(|m_{e}|^{2}, |m_{\mu}|^{2}, |m_{\tau}|^{2}).
\]  
The three eigenvalues of the matrix $M_{l}$, corresponding to the masses of the three charged leptons, satisfy the following relations.
\begin{equation}\label{W4q1}
     \text{Tr}|M_{L}M^{\dagger}_{L}|=|m_{e}|^{2}+|m_{\mu}|^{2}+|m_{\tau}|^{2}
 \end{equation}
 \begin{equation}\label{W4q2}
     \text{Det}|M_{L}M^{\dagger}_{L}|=|m_{e}|^{2}|m_{\mu}|^{2}|m_{\tau}|^{2}
 \end{equation}
 \begin{equation}\label{W4q3}
     \Big(\text{Tr}|M_{L}M^{\dagger}_{L}|\Big)^{2}-\text{Tr}|(M_{L}M^{\dagger}_{L})|= 2\Big(|m_{e}|^{2}|m_{\mu}|^{2}+|m_{\mu}|^{2}|m_{\tau}|^{2}+|m_{e}|^{2}|m_{\tau}|^{2}\Big)
 \end{equation}
We have fixed the values of the free parameters ($a_{1}, a_{2}, a_{3}$) by using equations.~\eqref{W4q1}, \eqref{W4q2}, and \eqref{W4q3}.
%%%%%%%%%%%%%%%%%%%%%%%%%%%%%%%%%%%%%%%%%%%%%%%%%%%%%%%%%%%
\subsection{Yukawa Lagrangian for the neutral leptons}
The Yukawa Lagrangian for the neutral lepton sector contains the active neutrino mass term $\mathcal{L}_{L}$, the Dirac mass term $\mathcal{L}_{D}$, the Majorana mass term $\mathcal{L}_{N}$, and the right-handed–sterile neutrino mixing term $\mathcal{L}_{NS}$. The complete Lagrangian is given in equation ~\eqref{W4Q2}.\\
%%%%%%%%%%%%%%%%%%%%%%%%%%%%%%%%%
\begin{align}\label{W4Q2}
\mathcal{L} =\ & 
b_{1}\Delta_{L} (\overline{l^{C}_{L}} Y^{(0)}_{3})_{3_S}l_{L} 
+ b_{2}\Delta_{L}Y^{(0)}_{1}(\overline{l^{C}_{L}}l_{L}) 
+ b_{3} \Phi (l_{L}Y^{(0)}_{3})_{3_S} N_{R} \nonumber \\
+& b_{4} \Phi (l_{L}Y^{(0)}_{3})_{3_A} N_{R}
 + b_{5} \Phi (l_{L}N_{R}) Y^{(0)}_{1} 
+ b_{6}\Delta_{R}(\overline{N^{C}_{R}}Y^{(0)}_{3})_{3_S}N_{R}  \nonumber \\
+& b_{7}\Delta_{R}(\overline{N^{c}_{R}}N_{R})Y^{(0)}_{1}
 + b_{8}\chi_{R} (\overline{N_{R}}Y^{(0)}_{3})_{1}S  
+ b_{9}\chi_{R} (\overline{N_{R}}Y^{(0)}_{3})_{1^{\prime\prime}}S   \nonumber \\
+& b_{10}\chi_{R} (\overline{N_{R}}Y^{(0)}_{3})_{1^{\prime}}S
\end{align}
From the above Lagrangian, we have constructed the active neutrino mass matrix $M_{L}$, the Dirac mass matrix $M_{D}$, the Majorana mass matrix $M_{R}$, and the sterile–right-handed mixing matrix $M$, whose structures are provided below.
 \begin{equation}
     M_{L}=v_{L}\begin{pmatrix}
         2b_{1}Y^{(0)}_{3,1}+b_{2} & -b_{1}Y^{(0)}_{3,3} & -b_{1} Y^{(0)}_{3,2} \\
         -b_{1}Y^{(0)}_{3,3} & 2b_{1}Y^{(0)}_{3,2} & -b_{1}Y^{(0)}_{3,1}+b_{2}\\
         -b_{1}Y^{(0)}_{3,2} & -b_{1}Y^{(0)}_{3,1}+b_{2} & 2b_{1}Y^{(0)}_{3,3}
     \end{pmatrix}
 \end{equation}
 %%%%%%%%%%%%%%%%%%%%%%%%%%%%%%%%%%%%%
 \begin{equation}
     M_{D}= v\begin{pmatrix}
        2 b_{3} Y^{(0)}_{3,1}+b_{5} & (b_{3}-b_{4}) Y^{(0)}_{3,3} & (-b_{3}+b_{4} )Y^{0}_{3,2} \\
         (-b_{3}+b_{4}) Y^{(0)}_{3,3} & 2 b_{3} Y^{(0)}_{3,2} & (-b_{3}-b_{4}) Y^{0}_{3,1}+b_{5} \\
          (-b_{3}-b_{4}) Y^{(0)}_{3,2} & (-b_{1}+b_{4}) Y^{(0)}_{3,1}+b_{5} & 2 b_{2} Y^{(0)}_{3,3} 
     \end{pmatrix}
 \end{equation}
 %%%%%%%%%%%%%%%%%%%%%%%%%%%%%
  \begin{equation}
     M_{R}=\frac{v_{L}}{v_{R}}\begin{pmatrix}
         2b_{6}Y^{(0)}_{3,1}+b_{7} & -b_{6}Y^{(0)}_{3,3} & -b_{6} Y^{(0)}_{3,2} \\
         -b_{6}Y^{(0)}_{3,3} & 2b_{6}Y^{(0)}_{3,2} & -b_{6}Y^{(0)}_{3,1}+b_{7}\\
         -b_{6}Y^{(0)}_{3,2} & -b_{6}Y^{(0)}_{3,1}+b_{7} & 2b_{6}Y^{(0)}_{3,3}
     \end{pmatrix}
 \end{equation}
 %%%%%%%%%%%%%%%%%%%%%%%%%%%%%%%%%
 \begin{equation}
     M= v^{\prime}\begin{pmatrix}
        b_{8} Y^{(0)}_{3,1} & b_{9} Y^{(0)}_{3,3} & b_{10} Y^{(0)}_{3,2} \\
         b_{8} Y^{(0)}_{3,3} & b_{9} Y^{(0)}_{3,2} & b_{10} Y^{(0)}_{3,1} \\
          b_{8} Y^{(0)}_{3,2} & b_{9} Y^{(0)}_{3,1} & b_{10} Y^{(0)}_{3,3} 
     \end{pmatrix}
 \end{equation}
 %%%%%%%%%%%%%%%%%%%%%%%%%%%%%%%%%
Where $v_{L}$ is the VEV of the scalar triplet $\Delta_{L}$, whose value can be at the sub-eV scale. We have considered its value to be $0.01~\text{eV}$. The parameter $v = 246~\text{GeV}$ denotes the VEV of the bidoublet. Similarly, $V_{R}$ and $v^{\prime}$ represent the VEVs of the right-handed scalar triplet $\Delta_{R}$ and the right-handed scalar doublet $\chi_{R}$, respectively, and we have taken their value to be $10~\text{TeV}$. \\
%%%%%%%%
While constructing the model, we have considered the polyharmonic Maa\ss{} forms of weight zero and level three. Therefore, we obtain a total of four polyharmonic Maa\ss{} forms. Among these, three transform as a triplet 
\[
Y^{(0)}_{3} = \left(Y^{(0)}_{3,1}, Y^{(0)}_{3,2}, Y^{(0)}_{3,3}\right),
\]
and one transforms as a singlet $Y^{(0)}_{1}$. The singlet $Y^{(0)}_{1}$ is a constant, which we have taken to be unity. The $q$-expansion of the triplet representation is given in Eq.~\eqref{W4A1}.

After constructing all the required mass matrices, we obtain the light neutrino mass matrix, the right-handed neutrino mass matrix, and the sterile neutrino mass matrix by using equation ~\eqref{W4Q6}. 
%%%%%%%%%%%%%%%%%%%%%%%%%%%%%%%%%%%%%%%%%%%%%%%%%%%%%%%%%%%%%%%%%%%%%%
\section{Numarical Analysis}\label{W4S3}
To calculate the Yukawa couplings, we have varied the parameter $\tau$ randomly within the fundamental domain, and the Yukawa coupling values are obtained using equation \eqref{W4A1}. After calculating the Yukawa couplings values, we fixed the free parameters associated with the charged lepton mass matrix using equation \eqref{W4q1}. \\ 

Since the charged lepton mass matrix is non-diagonal, the PMNS matrix takes the form $U_{PMNS} = U^{\dagger}_{l}U_{\nu}$. The $U_{PMNS}$ matrix is generated using the neutrino oscillation parameters given in Table \ref{W4T2}, while $U_{l}$ is constructed by calculating the eigenvectors of the charged lepton mass matrix. The light neutrino mass matrix $m_{\nu}$, which is constructed from the model, can also be expressed as 
\[
m_{\nu} = U_{L} U_{PMNS} m_{\text{diag}} (U_{L} U_{PMNS})^{T},
\]
where $m_{\text{diag}} = \text{diag}(m_{1}, m_{2}, m_{3})$. From neutrino oscillation data, we know the two mass-squared differences, the three mixing angles, and the Dirac CP-violating phase $\delta_{CP}$. Assuming the lightest neutrino mass lies within the range $10^{-5}$ to $0.1$ eV, the other two mass eigenvalues can be expressed in terms of the mass-squared differences as follows:

\begin{equation}
\begin{aligned}
    m_{\text{diag}} &= \left(m_1, \sqrt{m_1^2 + \Delta m^2_{21}}, \sqrt{m_1^2 + \Delta m^2_{31}}\right), \quad &\text{(for NH)} \\
    m_{\text{diag}} &= \left(\sqrt{m_3^2 + \Delta m^2_{23} - \Delta m^2_{21}}, \sqrt{m_3^2 + \Delta m^2_{23}}, m_3\right), \quad &\text{(for IH)}
\end{aligned}
\end{equation} 

Finally, the unknown free parameters $b_{1}$ and $b_{2}$ are determined using the $3\sigma$ values of neutrino oscillation parameters together with the Yukawa couplings ($Y^{(0)}_{3,1},\,Y^{(0)}_{3,2},\,Y^{(0)}_{3,3}$) obtained by randomly varying $\tau$ in the upper half of the complex plane. The obtained maximum and minimum value of the three Yukawa couplings is given in the Table \ref{W4T3}.

\begin{figure}[H]
    \centering
    \begin{subfigure}[b]{\columnwidth}
        \centering
        \includegraphics[width=\linewidth]{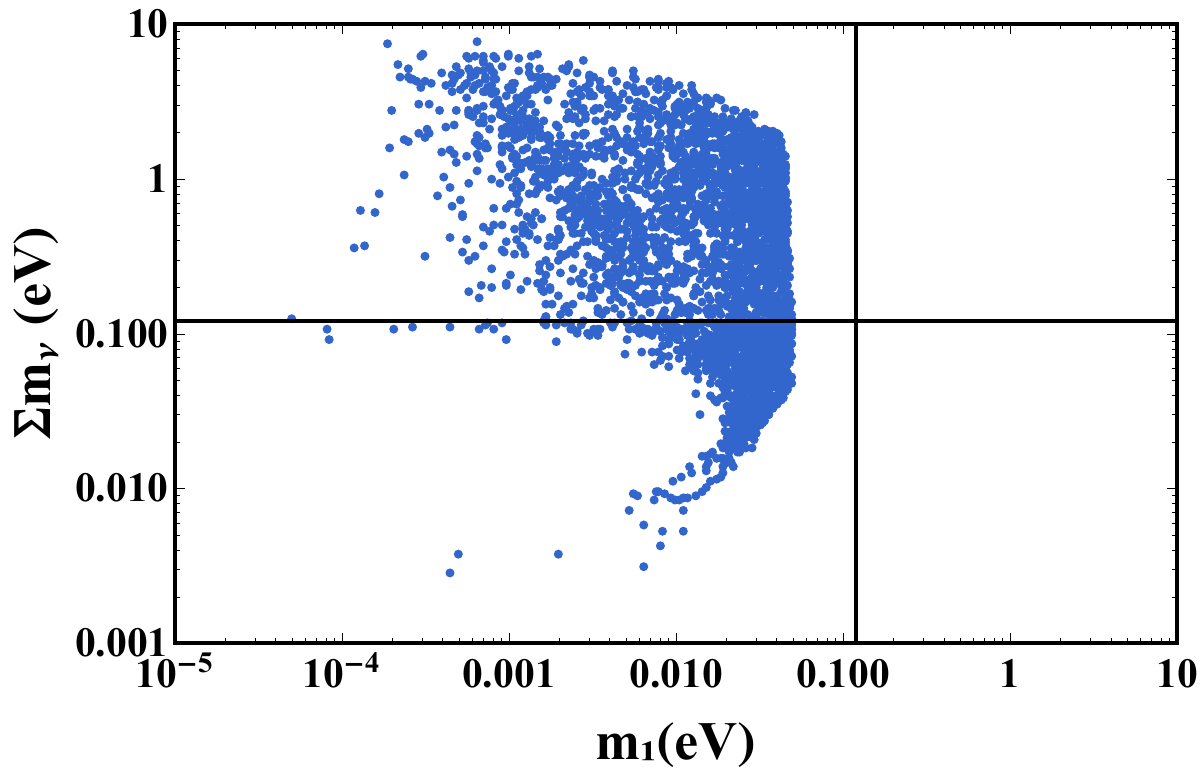}
        \caption{NH}
        \label{w4f1}
    \end{subfigure}
    \vskip\baselineskip
    \begin{subfigure}[b]{\columnwidth}
        \centering
        \includegraphics[width=\linewidth]{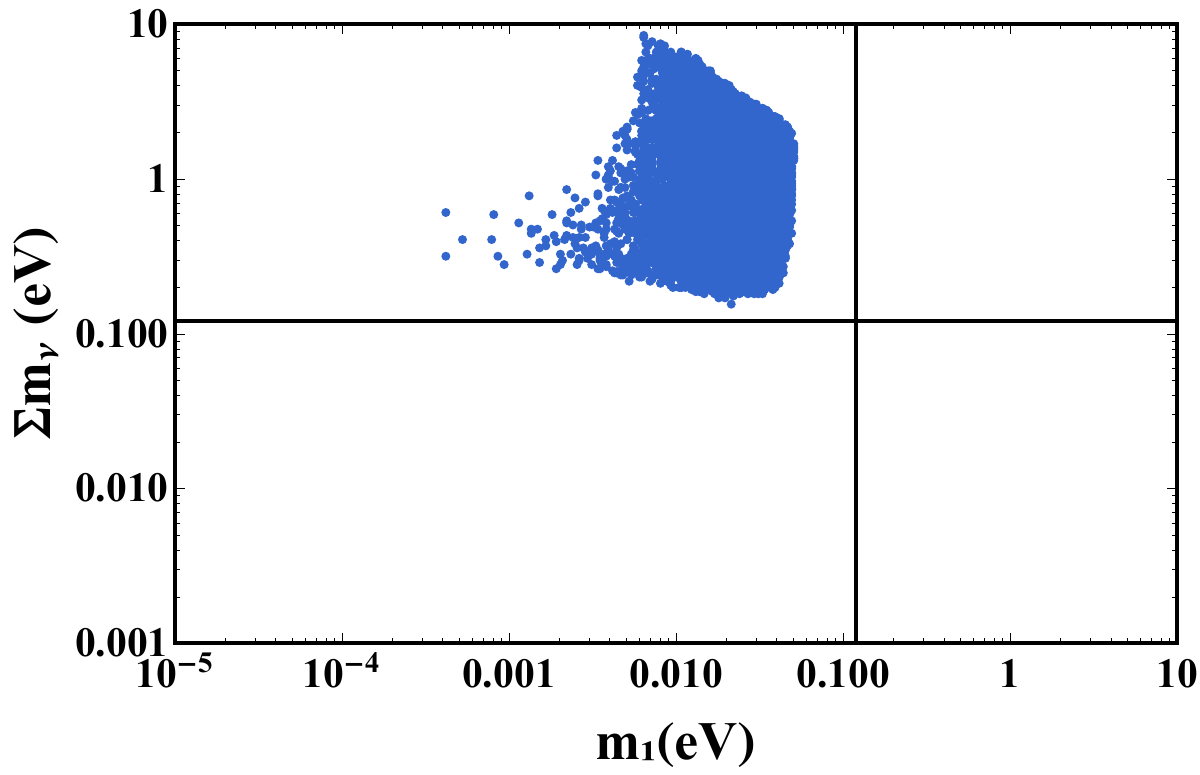}
        \caption{IH}
        \label{w4f2}
    \end{subfigure}
    \caption{Variation of the sum of the neutrino mass with the lightest neutrino mass.}
    \label{W4F2}
\end{figure}

\begin{figure}[H]
    \centering
    \begin{subfigure}[b]{\columnwidth}
        \centering
        \includegraphics[width=\linewidth]{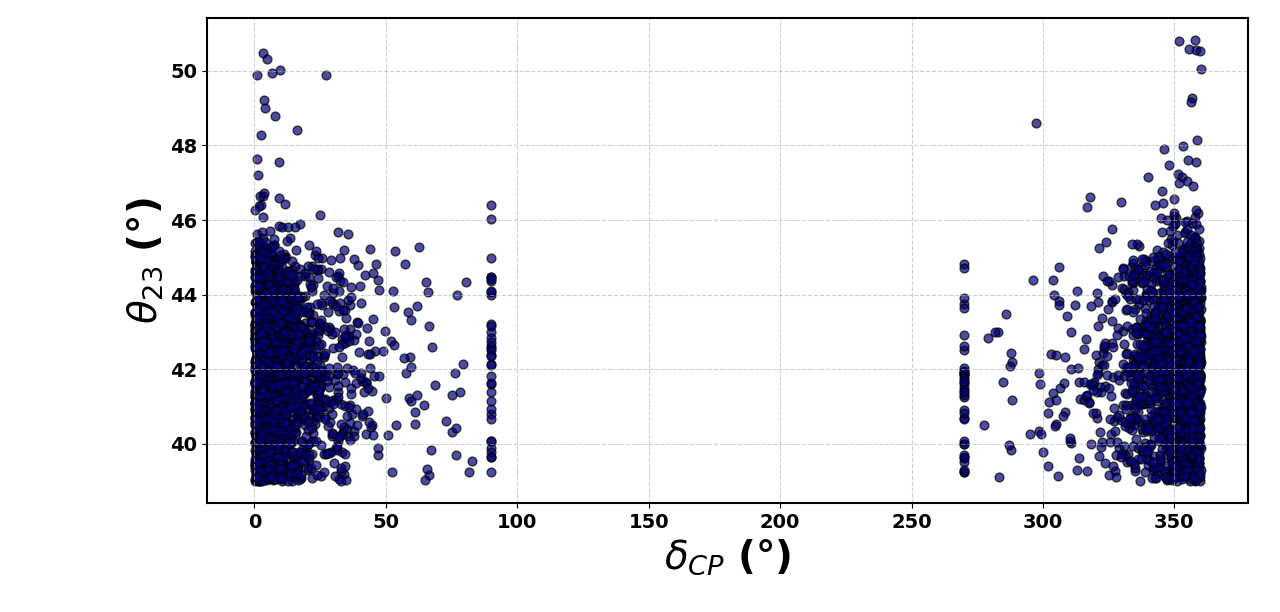}
    \end{subfigure}
    \caption{Parameter space of of $\theta_{23}$ and $\delta_{CP}$.}
    \label{W4F3}
\end{figure}

\begin{figure}[H]
    \centering
    \begin{subfigure}[b]{\columnwidth}
        \centering
        \includegraphics[width=\linewidth]{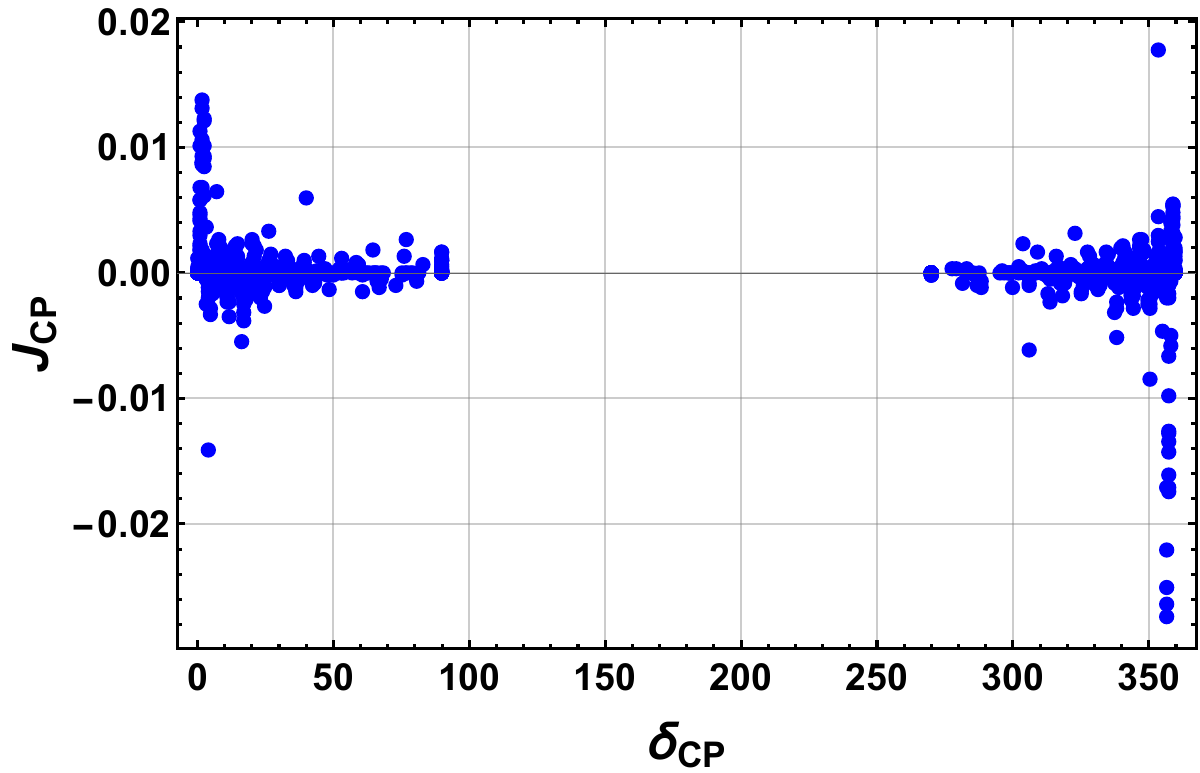}
    \end{subfigure}
    \caption{Variation of the Jarlskog invariant parameter with the Dirac CP phase.}
    \label{W40F3}
\end{figure}

\begin{table}[ht]
    \centering
    \begin{tabular}{|c|c|c|}
        \hline
        $|Y|$ & min & max \\ \hline
        $|Y^{(0)}_{3,1}|$ & 0.77432 & 0.792296 \\ \hline
        $|Y^{(0)}_{3,2}|$ & 0.00762 & 0.229353 \\ \hline
        $|Y^{(0)}_{3,3}|$ &  0.00006 & 0.05496\\ \hline
    \end{tabular}
    \caption{Parameter space of Yukawa couplings}
    \label{W4T3}
\end{table}
%%%%%%%%%%%%%%%%%%%%%%%%%%%%%%%%%%%%%%%%%%%%%%%%%%%%%%%%%%%%%
\begin{figure}[H]
    \centering
    \begin{subfigure}[b]{\columnwidth}
        \centering
        \includegraphics[width=\linewidth]{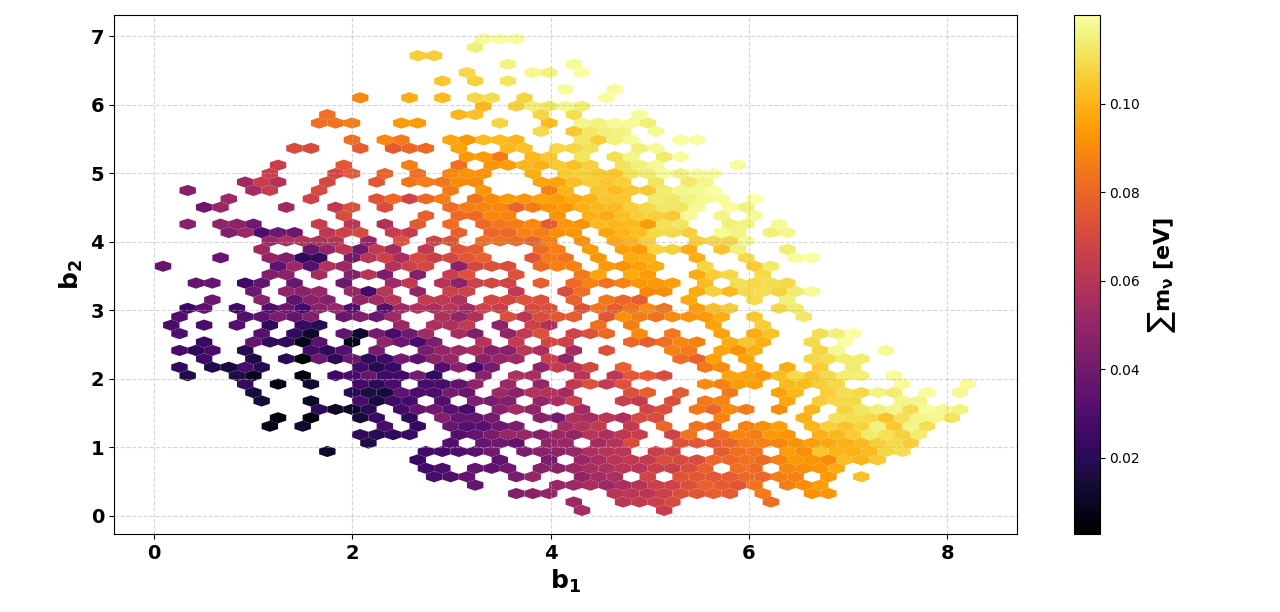}
    \end{subfigure}
    \caption{Parameter space of absolute value of $b_{1}$ and $b_{2}$ for NH.}
    \label{W4F1}
\end{figure}
%%%%%%%%%%%%%%%%%%%%%%%%%%%%%%%%%%%%%%%%%%%%%%%%%%
 After determining the free parameters, we calculated the sum of the neutrino masses $\sum m_{\nu}$, the mixing angle $\theta_{23}$, and the CP-violating phase $\Delta_{CP}$ from the model. We observed that the calculated value of $\sum m_{\nu}$ lies within the Planck bound (i.e., below $0.12$ eV) only for the NH case. For the IH scenario, no data points were found to satisfy the Planck bound. Figure \ref{W4F1} shows the parameter space of the free parameters $b_{1}$ and $b_{2}$ for which the sum of the neutrino mass is within the Planck bound. Figure \ref{W4F2} presents the predicted values of $\sum m_{\nu}$ and the lightest neutrino mass for both NH and IH. The results indicate that, for IH, the model fails to reproduce neutrino masses within the Planck limit and is unable to accommodate all neutrino oscillation parameters within the $3\sigma$ range.  Therefore, in the subsequent analysis of the effective Majorana mass and dark matter, we consider only the NH scenario. Since the octant of the mixing angle $\theta_{23}$ and the CP-violating phase $\delta_{CP}$ are still not precisely determined from neutrino oscillation experiments, we have calculated the corresponding values within the framework of our model. Figure \ref{W4F3} illustrates the parameter space of $\theta_{23}$ and $\delta_{CP}$ as obtained from the model. We have also calculated the $J_{\text{CP}}$ value using equation \eqref{W4Q8}. Figure \ref{W40F3} shows the parameter space of the calculated $J_{\text{CP}}$ values from the model. 

%%%%%%%%%%%%%%%%%%%%%%%%%%%%%%%%%%%%%%%%%%%%%%%%%%%%%%%
%----------------------------------------------------------
\begin{figure}[H]
    \centering
    \begin{minipage}{0.48\textwidth}
        \centering
        \includegraphics[width=\linewidth]{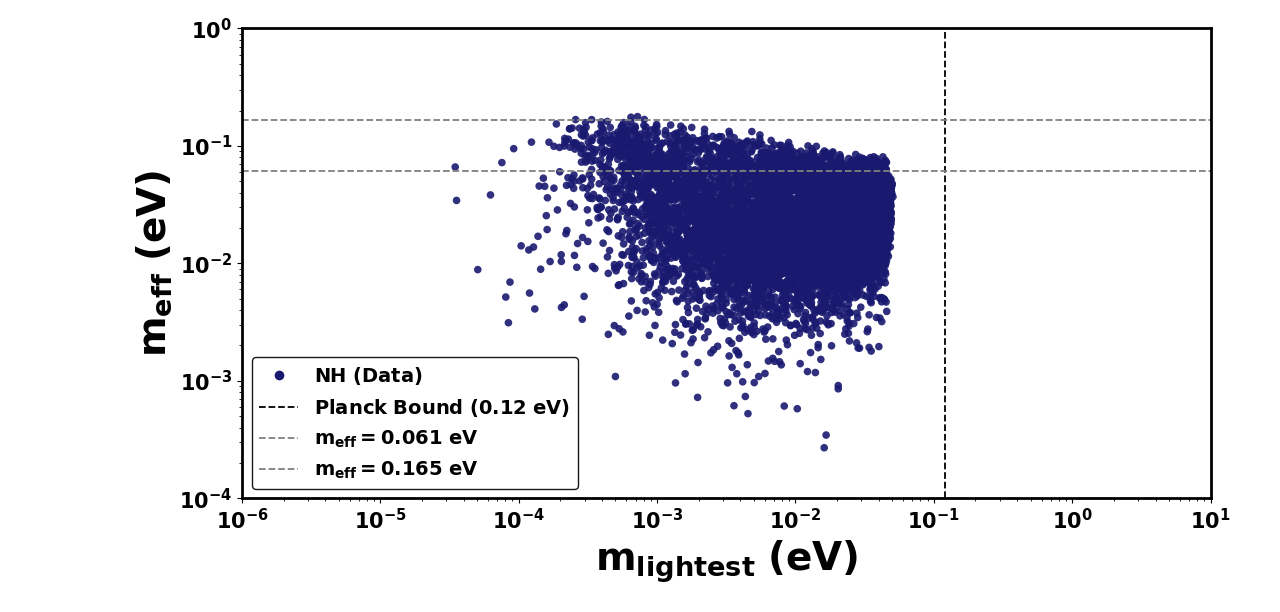}
        \caption{Variation of effective mass with the lightest neutrino mass.}
        \label{W4F4}
    \end{minipage}
\end{figure}

\begin{figure}
    \centering
    \begin{subfigure}[b]{\columnwidth}
        \centering
        \includegraphics[width=\linewidth]{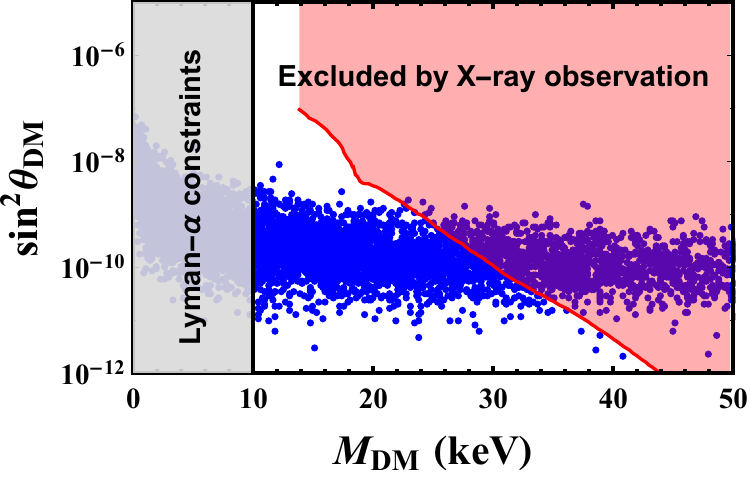}
        \caption{Variation of the active–sterile mixing angle with the dark matter mass.}
        \label{w4f3}
    \end{subfigure}
    \vskip\baselineskip
    \begin{subfigure}[b]{\columnwidth}
        \centering
        \includegraphics[width=\linewidth]{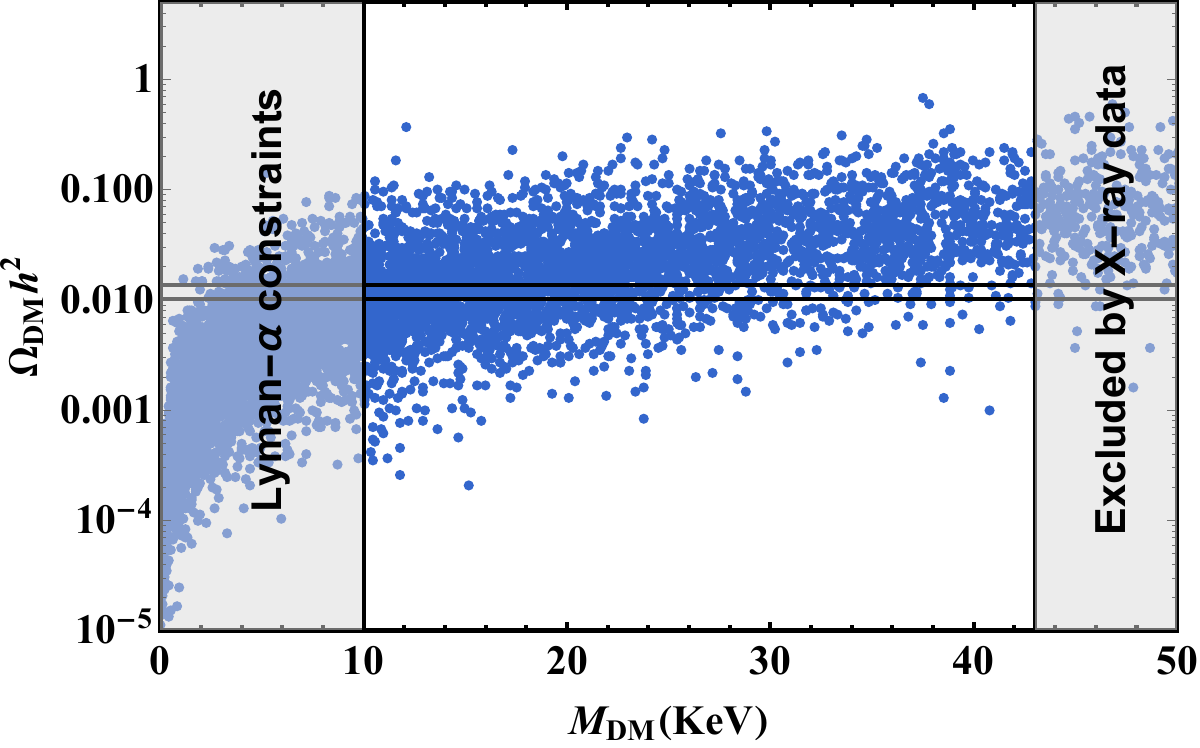}
        \caption{Variation of the relic density with the dark matter mass.}
        \label{w4f4}
    \end{subfigure}
     \vskip\baselineskip
    \begin{subfigure}[b]{\columnwidth}
        \centering
        \includegraphics[width=\linewidth]{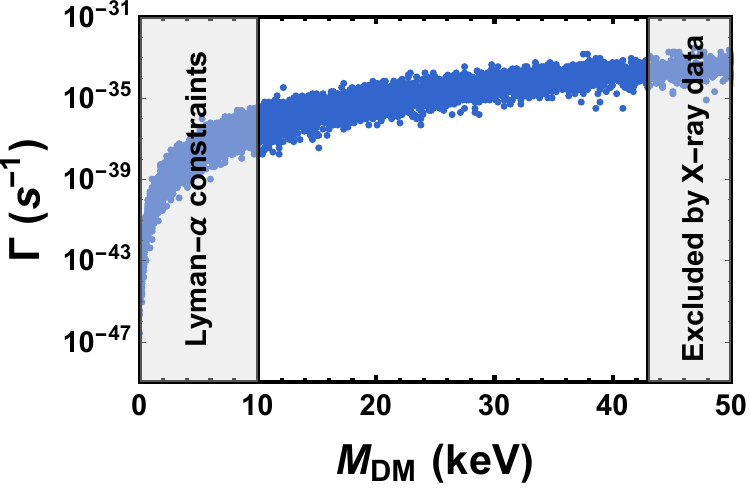}
        \caption{Variation of the decay width of the dark matter particle into SM particles as a function of the dark matter mass}
        \label{w4f5}
    \end{subfigure}
    \caption{Variation of the active–sterile mixing angle, relic density, and decay width with the mass of the dark matter candidate}
    \label{W4F5}
\end{figure}

%%%%%%%%%%%%%%%%%%%%%%%%%%%%%%%%%%%%%%%%%%%%%%%%%%%%%%%%%%%%%%%%%%%%%%%%%%%%%%%%%%%%%%%%%%%%%%%%%%%%%%%%%%%%%%%%%%%%%%%%%%%
We have also studied neutrinoless double beta decay ($0\nu\beta\beta$) and calculated the effective Majorana mass arising from the standard contribution. The calculated values are plotted against the lightest neutrino mass, as shown in Figure \ref{W4F4}. In this figure, the two horizontal lines represent the experimental upper bounds on the effective mass, while the vertical line indicates the Planck bound on the $\sum m_{\nu}$.
%%%%%%%%%%%%%%%%%%%%%%%%%%%%%%%%%%%%%%%%%%%%%%%%%%%%%%%%%%%%%%%%
We have also explored the dark matter sector of the model, considering only the NH scenario. The model contains sterile neutrinos, which can serve as potential dark matter candidates. As discussed in \cite{Drewes:2016upu}, keV-scale sterile neutrinos are viable dark matter candidates, and the sterile neutrino in the present model also lies in the keV mass range. We have calculated the relic abundance and the decay width of the sterile neutrino into SM neutrinos using Eqs.~\eqref{W4Q2} and \eqref{W4Q3}, respectively. These calculated quantities are then plotted against the mass of the dark matter candidate, identified as the lightest sterile neutrino in the model. In all the figures presented in Fig.~\ref{W4F5}, the vertical line at $10\,\text{keV}$ denotes the Lyman-$\alpha$ constraint on the dark matter mass, while the shaded region in Fig.~\ref{w4f3} corresponds to the X-ray constraints.
%-------------------------------------------------
%--------------------------------------------------------------------
\section{Results and Discussion}\label{W4S4}
The model includes four Yukawa couplings, $Y^{(0)}_{3,1}$, $Y^{(0)}_{3,2}$, and $Y^{(0)}_{3,3}$, each having both holomorphic and non-holomorphic components and $Y^{(0)}_{1}$ which is a constant and we have considered it as one. The absolute values of these couplings are found to range from approximately $0.79$ down to the order of $10^{-5}$. Using these Yukawa coupling values along with the $3\sigma$ ranges of the neutrino oscillation parameters, the free parameters $b_{1}$ and $b_{2}$ are determined.
It is observed that for the NH, the sum of the neutrino masses lies within the Planck bound when the absolute values of the free parameters $b_{1}$ and $b_{2}$ vary approximately between $1$ and $8$. In contrast, for the IH, no parameter region of $b_{1}$ and $b_{2}$ is found that yields a sum of neutrino masses consistent with the Planck bound.\\
From our analysis, we find that the model is favourable towards the NH scenario, as it successfully reproduces all the neutrino oscillation parameters only in this case. Most of the calculated values of the sum of neutrino masses lie within the Planck bound, further supporting the viability of the model. In addition, the model shows a clear preference for the lower octant of the atmospheric mixing angle $\theta_{23}$, with nearly all the predicted values lying in the range $40^\circ$--$45^\circ$. The model also provides specific predictions for the Dirac CP-violating phase $\delta_{CP}$, which lies in the ranges $[0^\circ, 50^\circ]$ and $[280^\circ, 350^\circ]$, while no solutions are obtained in the interval $[50^\circ, 250^\circ]$ also the calculated value of $J_{\text{CP}}$ lie in the range of $-0.0284$ to $0.0177$. Furthermore, the effective Majorana mass is predicted to vary from $0.1~\text{eV}$ down to $10^{-3}~\text{eV}$ as the lightest neutrino mass decreases from $0.12~\text{eV}$ to $10^{-4}~\text{eV}$. Importantly, the predicted values of the effective Majorana mass remain consistent with the current upper experimental bounds. Since the model accommodates a keV-scale sterile neutrino, we also investigate its role as a potential dark matter candidate. After imposing the cosmological constraints, the allowed parameter space for the dark matter mass is found to lie in the range $10~\text{keV}$--$30~\text{keV}$. We further calculate the relic abundance and observe that, within this mass range, a significant number of parameter points remain consistent with the observed relic abundance. Additionally, the decay rate of the sterile neutrino dark matter candidate into SM neutrinos is obtained in the range $10^{-35}$--$10^{-34}~\text{sec}^{-1}$ for dark matter masses between $10$ and $30~\text{keV}$.
%%%%%%%%%%%%%%%%%%%%%%%%%%%%%%%%%%%%%%%%%%%%%%%%%%%%%%%%%%%%%%%%%%%%%%%%%%%%%%%%%%%%%%%%%%%%%%%%%%%%%5
\section{Conclusion}\label{W4S5}

In this work, we have realized a non-supersymmetric left right model using the $\Gamma_{3}$ modular group. Since the framework is non-supersymmetric, we have implemented the idea of non-holomorphic modular symmetry, where the Yukawa couplings are described by polyharmonic Maa\ss{} forms. A notable advantage of employing non-holomorphic modular symmetry is that it allows Yukawa couplings with modular weight zero as well as negative weights, which are not permitted in conventional modular symmetry.  In our construction, the Yukawa couplings are assigned modular weight zero, consistent with the choice of zero modular weight for all fermion and scalar fields in the model. Nevertheless, the Yukawa couplings retain explicit $\tau$-dependence due to their formulation in terms of polyharmonic Maa\ss{} forms. This is in sharp contrast to the conventional modular symmetry framework, where a zero modular weight assignment would simply lead to constant matrices.  \\
Phenomenologically, the model shows a clear preference for the normal hierarchy of neutrino masses and predicts the atmospheric mixing angle $\theta_{23}$ in the lower octant. Since both the neutrino mass ordering and the octant of $\theta_{23}$ remain unresolved in current experiments, future neutrino oscillation data favoring NH and the lower octant of $\theta_{23}$ would provide strong support for our framework. In addition, the model accommodates a keV-scale sterile neutrino as a dark matter candidate. The allowed mass range for the sterile neutrino is found to lie between $10~\text{keV}$ and $30~\text{keV}$, consistent with the observed relic abundance of dark matter. Future neutrino oscillation and dark matter searches will play a crucial role in testing the predictions of this model. \\ 

Overall, the presented framework successfully addresses neutrino mass and mixing as well as dark matter, while requiring only a minimal particle content. Importantly, it avoids the introduction of flavons and does not rely on supersymmetry, which has not yet been experimentally established. Therefore, exploring such non-supersymmetric realizations of modular symmetry provides a promising direction for model building beyond the SM.  
%%%%%%%%%%%%%%%%%%%%%%%%%%%%%%%%%%%%%%%%%%%%%%%%%%%%%%%%%%%%%%%%%%%%%%%%%%%%%%%%%%%%%%%%%%%%%%%%%%%%%%%%%%%%%%%%%%%%%%%%%%%%%%%%%%%%%%%%%%%%%%%%%%%%
\appendix
\section{\textbf{Modular Group and Polyharmonic $Maa\beta$ form}} \label{W2A1}
The modular group $SL(2,Z)=\Gamma$ is defined as a group of $2\times2$ matrices with positive or negative integer element and determinant equal to $1$ and it represents the symmetry of a torus \cite{deAnda:2023udh,deAdelhartToorop:2011re}. It is infinite group and generated by two generators of the group $S$ and $T$.\\
	\begin{equation}\label{Q1}
		\Gamma = \Biggl\{\begin{pmatrix}
			a & b\\
			c & d
		\end{pmatrix} | a,b,c,d \in \mathbb{Z}, ad-bc=1\Biggr\}~~ .
	\end{equation}
	The generator of the group satisfy the conditions:\\
	$S^{2} = 1$ \hspace{0.5cm} and  \hspace{0.5cm} $(ST)^{3} = 1$ \\
	and they can be represented by $2\times 2$ matrices 
	%%%%%%%%%%%%%%%%%%%%%%%%%%%%%%%%%%%%%%55
	\begin{align*}
		S=\begin{pmatrix} 
			0 & 1\\
			-1 & 0 \\
		\end{pmatrix},
		&  ~~~~~~
		T= \begin{pmatrix}
			1 & 1 \\
			0 & 1 \\
		\end{pmatrix} ~~~.
	\end{align*}
	%%%%%%%%%%%%%%%%%%%%%%%%%%%%%%%%%%%%
	A two-dimensional space is obtained, when the torus is cut open and this two-dimensional space can be viewed as an Argand plane and modulus $\tau$ is the lattice vector of that Argand plane. The transformation of modulus $\tau$ \cite{Ferrara:1989qb} of the modular group on the upper half of the complex plane is given below 
	\begin{equation*}
		\gamma : \tau \rightarrow \gamma(\tau) = \frac{(a\tau + b)} {(c\tau + d)}
	\end{equation*} 
	%%%%%%%%%%%%%%%%%%%%%%%%%%%%%%%%%%%%%%%%%%%%%%%%%%%%%%%%%%%%%%
	the transformation of $\tau$ is same for both $\gamma$ and $-\gamma$ and we can define a group $\bar{\Gamma}= PSL(2,Z)$, which is a  projective special linear group. Also, the modular group has an infinite number of normal subgroups, which is the principal congruence subgroup of level N and can be defined as
	\begin{equation}\label{Q2}
		\Gamma(N)= \Biggl\{\begin{pmatrix}
			a & b\\
			c & d
		\end{pmatrix} \in SL(2,Z), \begin{pmatrix}
			a & b \\
			c & d
		\end{pmatrix}= \begin{pmatrix}
			1 & 0 \\
			0 & 1
		\end{pmatrix} (mod N)\Biggr\},
	\end{equation}  
	%%%%%%%%%%%%%%%%%%%%%%%%%%%%%%%%%%%%%%%%%
	and for $N>2$, $\bar{\Gamma}(N)=\Gamma_{N}$. In model building purpose it is efficient to use finite group. Usually a modular group is an infinite group but we can obtain a finite modular group for $N>2$, if we consider the quotient group $\Gamma_{N} =PSL(2,Z)/\bar{\Gamma}(N)$ and those modular group are isomorphic to non-abelian discrete groups. \\
	The modular invariance required the Yukawa couplings to be a certain modular function $Y(\tau)$ and should fellow the following transformation property.
	
	\begin{equation}\label{n2}
		Y(\gamma\tau)=(c\tau+d)^{k}Y(\tau)
	\end{equation} 
	%%%%%%%%%%%%%%%%%%%%%%%%%%%%%%%%%%%%%%%%%%%%%
	One can realise the non-supersymmetric framework by using the framework of automorphic form and the assumption of holomorphicity is replaced by the Laplacian condition and in such case the Yukawa coupling can have both holomorphic and non-holomorphic parts \cite{Qu:2024rns,Ding:2020zxw,Ding:2024inn}. In the present work, we are concerned with the polyharmonic $Maa\beta$ forms of weight k and the Yukawa coupling needs to follow another transformation property, which is given below
	%%%%%%%%%%%%%%%%%%%%%%%%%%%%%%%%%%%%%%%%
	\begin{equation}\label{n3}
		\Delta_{k}Y(\tau)=0
	\end{equation}
	%%%%%%%%%%%%%%%%%%%%%%%%%%%%%%%%%%%%%%%%%%%%%%%% 
	where $\tau=x+iy$ and $\Delta_{k}$ is the hyperbolic Laplacian operator
	\begin{equation}
		\Delta_{k}=-y^{2}\big(\frac{\partial^{2}}{\partial x^{2}} + \frac{\partial^{2}}{\partial y^{2}}\big) + iky\big(\frac{\partial}{\partial x}+ i\frac{\partial}{\partial y}\big) = -4y^{2} \frac{\partial}{\partial \tau}\frac{\partial}{\partial \tau}+ 2iky \frac{\partial}{\partial \tau}
	\end{equation}
	%%%%%%%%%%%%%%%%%%%%%%%%%%%%%%%%%%%%%%%%%%%%%%%%%%%%%%
	The weight k of polyharmonic $Maa\beta$ forms can be positive, zero, or even negative. Based on the transformation property given in the equation \eqref{n2}, which implies that $Y(\tau + N)=Y(\tau)$ and considering the transformation property from equation \eqref{n3}, the Fourier expansion of a level N and weight k polyharmonic $Maa\beta$ form can be expressed as\cite{Qu:2024rns}
	%%%%%%%%%%%%%%%%%%%%%%%%%%%%%%%%%%%%%%%%%%%%%%
	\begin{equation}\label{n01}
		Y(\tau)= \sum_{n\in\frac{1}{N}\mathbb{Z} n\leqq 0} c^{+}(n)q^{n} + c^{-}(0)y^{1-k} + \sum_{n\in \frac{1}{N}\mathbb{Z}n\leq 0} c^{-}(n)\Gamma(1-k,-4\pi ny)q^{n}
	\end{equation}
	%%%%%%%%%%%%%%%%%%%%%%%%%%%%%%%%%%%%
	where $q=e^{i2\pi\tau}$
	%%%%%%%%%%%%%%%%%%%%%%%%%%%%%%%%%%
	In the present work, since we are working in the non-supersymmetric framework, we have focused on polyharmonic $Maa\beta$ forms of weight zero at level $3$. It can be arranged into a singlet and triplet under $A_{4}$ group. The q expansion of the weight zero Yukawa couplings at level three is provided below

\begin{align}\label{W4A1}
Y^{(0)}_{3,1} &= y - \frac{3 e^{-4 \pi y}}{\pi q}
- \frac{9 e^{-8\pi y}}{2\pi q^{2}}
+ \frac{12\pi y}{\pi q^{3}}
- \frac{21 e^{-16\pi y}}{4\pi q^{4}} \notag \\
&\quad - \frac{18 e^{-20\pi y}}{5\pi q^{5}}
- \frac{3e^{-24\pi y}}{2\pi q^{6}}
+ \cdots \notag \\
&\quad - \frac{9\log 3}{4\pi}
- \frac{3q}{\pi}
- \frac{9q^{2}}{2\pi}
- \frac{q^{3}}{\pi}
- \frac{21q^{4}}{4\pi}
- \frac{18q^{5}}{5\pi}
- \frac{3q^{6}}{2\pi} , \\[6pt]
%%%%%%%%%%%%%%%%%%%%%%%%%%%%%%%%%%%%%%%%%%%%%%%%%%%%%%%%%%
Y^{(0)}_{3,2} &= \frac{27q^{1/3}e^{\pi y/3}}{\pi}
\Big( \frac{e^{-3\pi y}}{4q}
+ \frac{e^{-7\pi y}}{5q^{2}}
+ \frac{5e^{-11\pi y}}{16q^{3}} \notag \\
&\quad + \frac{2e^{-15\pi y}}{11q^{4}}
+ \frac{2e^{-19\pi y}}{7q^{5}}
+ \frac{4e^{-23\pi y}}{17 q^{6}}
+ \cdots \Big) \notag \\
&\quad + \frac{9q^{1/3}}{2\pi}
\Big( 1 + \frac{7q}{4} + \frac{8q^{2}}{7}
+ \frac{9q^{3}}{5}
+ \frac{14q^{4}}{13}
+ \frac{31q^{5}}{16}
+ \frac{20q^{6}}{19}
+ \cdots \Big) , \\[6pt]
%%%%%%%%%%%%%%%%%%%%%%%%%%%%%%%%%%%%%%%%%%%%%%%%%%%%%%%%%%
Y^{(0)}_{3,3} &= \frac{9q^{2/3}e^{2\pi y/3}}{2\pi}
\Big( \frac{e^{-2\pi y}}{q}
+ \frac{7e^{-6\pi y}}{4q^{2}}
+ \frac{8e^{-10\pi y}}{7q^{3}} \notag \\
&\quad + \frac{9e^{-14\pi y}}{5q^{4}}
+ \frac{14e^{-18\pi y}}{13q^{5}}
+ \frac{31e^{-22\pi y}}{16q^{6}}
+ \cdots \Big) \notag \\
&\quad + \frac{27q^{2/3}}{\pi}
\Big( \frac{1}{4} + \frac{q}{5}
+ \frac{5q^{2}}{16}
+ \frac{2q^{3}}{11}
+ \frac{2q^{4}}{7}
+ \frac{9q^{5}}{17}
+ \frac{21q^{6}}{20}
+ \cdots \Big) .
\end{align}
   
\bibliographystyle{abbrvnat} 
\bibliography{ref}

\end{document}